\documentclass[aps,prl,superscriptaddress,reprint]{revtex4-1}
\usepackage{amsmath}    
\usepackage{graphicx}   
\usepackage[dvipdfx,cmyk]{xcolor}     
\usepackage[caption=false]{subfig}  
\usepackage[colorlinks,linkcolor=red,citecolor=brown,urlcolor=blue]{hyperref}
\usepackage{dsfont,bm}
\usepackage{color}
\usepackage{pdfpages}

\usepackage{amsmath} 
\usepackage{amsthm} 
\usepackage{amssymb}	
\usepackage{graphicx} 
\renewcommand{\v}[1]{\ensuremath{\mathbf{#1}}} 

\newcommand{\ket}[1]{\left| #1 \right>} 
\let\baraccent=\= 
\renewcommand{\=}[1]{\stackrel{#1}{=}} 

\theoremstyle{definition}

\theoremstyle{remark}

\newcommand{\be}{\begin{equation}}
\newcommand{\ee}{\end{equation}}
\newcommand{\bA}{\begin{align}}
\newcommand{\eA}{\end{align}}
\newcommand{\bF}{\begin{figure}}
\newcommand{\eF}{\end{figure}}

\begin{document}
\title{Linear Optical Quantum Computing in a Single Spatial Mode}
\author{Peter C. Humphreys}
\author{Benjamin J. Metcalf}
\author{Justin B. Spring}
\author{Merritt Moore}
\affiliation{Clarendon Laboratory, Department of Physics, University of Oxford, OX1 3PU, United Kingdom}
\author{Xian-Min Jin}
\affiliation{Clarendon Laboratory, Department of Physics, University of Oxford, OX1 3PU, United Kingdom}
\affiliation{Department of Physics, Shanghai Jiao Tong University, Shanghai 200240, PR China}
\author{Marco Barbieri}
\author{W. Steven Kolthammer}
\author{Ian A. Walmsley}
\affiliation{Clarendon Laboratory, Department of Physics, University of Oxford, OX1 3PU, United Kingdom}

\begin{abstract}
We present a scheme for linear optical quantum computing using time-bin encoded qubits in a single spatial mode. We show methods for single-qubit operations and heralded controlled phase (CPhase) gates, providing a sufficient set of operations for universal quantum computing with the Knill-Laflamme-Milburn~\cite{Knill2001} scheme. Our protocol is suited to currently available photonic devices and ideally allows arbitrary numbers of qubits to be encoded in the same spatial mode, demonstrating the potential for time-frequency modes to dramatically increase the quantum information capacity of fixed spatial resources. As a test of our scheme, we demonstrate the first entirely single spatial mode implementation of a two-qubit quantum gate and show its operation with an average fidelity of $0.84\pm0.07$.
\end{abstract}
\maketitle

\emph{Introduction-} Linear optics provides a promising platform for universal quantum computing~\cite{Knill2001, Kok2007,Ralph2010209}. Although logical gates can only be implemented probabilistically, Knill, Laflamme, and Milburn (KLM) have shown that they can be rendered deterministic by making use of ancillary resources, measurements and feed-forward~\cite{Knill2001}. However, the overhead is large, and this presents one of the most significant challenges to the scalability of all proposed linear-optical quantum computing (LOQC) implementations~\cite{Kok2007,Ralph2010209}. To date, demonstrations of experimental schemes have mainly adopted spatial degrees of freedom for the manipulation of quantum states~\cite{Pittman2001a, Gasparoni2004, O'Brien2003, Bao2007a, Lanyon2008, Okamoto2011, Ralph2010209, Kok2007}. Consequently, scalable implementations of even few-qubit protocols in LOQC demand many spatial modes and complex routing networks with active switches, necessary to implement feed-forward~\cite{Ma2011}.  

Modern telecommunication suggests a promising alternative or complement to spatial schemes in its extensive use of time-frequency encodings. The same approach for quantum information and communication protocols naturally provides access to high dimensional Hilbert spaces~\cite{DeRiedmatten2004, Smith2010, Nisbet-jones2013} while maintaining a compact device design, and can leverage the existing classical communications technology base. Additionally, time-frequency encodings benefit from a relative insensitivity to inhomogeneities in transmission mediums~\cite{Thew2002, Smith2010}. These advantages have been recognized in works exploring the preparation of time-frequency entangled states~\cite{Durt2001, Simon2005, Barreiro2005,Zavatta2006}, including their use in the violation of Bell inequalities~\cite{Franson1989, Olislager2010}, quantum key distribution~\cite{Tittel2000}, teleportation~\cite{Marcikic2003}, and continuous-variable cluster states~\cite{Menicucci2011}.

Quantum computing based on time-frequency encoding has received comparatively little attention, but has become increasingly feasible with the advent of fast switchable integrated phase controllers ~\cite{Hall2011, Bonneau2012}. This was highlighted by a recent classical simulation of a quantum random walk based on a time-bin encoding and fast polarization switching~\cite{Schreiber2012}. Previous studies have explored unitary operations for time~\cite{Soudagar2007,Bussieres2006} and frequency encodings~\cite{Huntington2004}, but these implementations have relied on conversion from time-frequency to multiple spatial modes for manipulation. 

Here we present a concept for linear optical quantum computing using time-bin encoded qubits and only a single spatial mode. Time bins provide a practical solution for the manipulation and detection of time-frequency modes with current technology. We outline methods that provide a sufficient set of operations to allow for universal quantum computing with the KLM scheme. In order to illustrate our scheme, we demonstrate experimentally the first implementation of a two-qubit quantum gate in a single spatial mode and show its high fidelity of operation.\\

\emph{Scheme-} We consider a string of time-bin encoded qubits in a single spatial mode. The polarization degree of freedom is used to define a `register' polarization, in which qubits are stored and transmitted, and a `processing' polarization in which specific time bins are briefly manipulated. After each processing stage, all qubits are returned to the register polarization to ensure that a high degree of coherence is maintained between the time bins during further transmission. 

\begin{figure}[htbp]
\centering
\includegraphics[width=8cm]{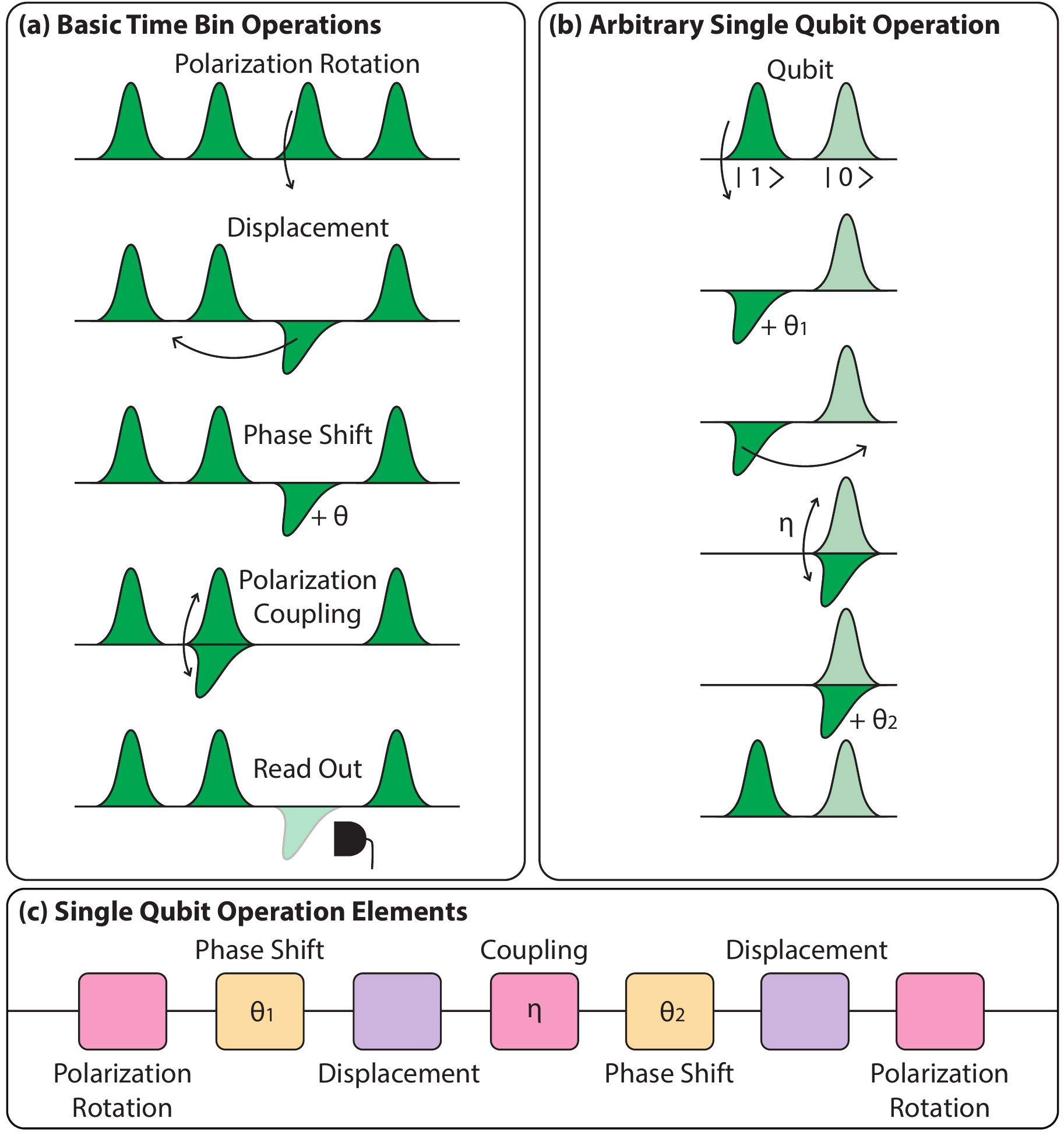}
\caption{(a) Complete set of basic operations necessary for the manipulation of a string of time bins in a single spatial mode. States are initially temporally encoded in the register polarization, shown as vertical. The first operation rotates a time bin to the horizontal processing polarization in order to enable subsequent manipulations as required. After manipulation, the time bins are rotated back into the register polarization in order to protect against dephasing. (b) Operations sufficient for arbitrary single-qubit operations. For brevity, the final displacement and rotation are implicit in the last line. (c) The minimal set of elements required to implement these single-qubit operations.}
\label{fig:scheme}
\end{figure}

Five basic operations are needed for our implementation, as shown in Fig.~\ref{fig:scheme}: a polarization rotation moves a time bin between register and processing polarizations; a displacement operation moves a time bin in the processing polarization forwards and backwards relative to time bins in the register polarization; a phase shift adds a specified phase between two polarizations; a polarization coupling operation is a partial polarization rotation between two orthogonally polarized time bins; and finally, a read out operation measures the number of photons in a specified bin. With the exception of read out, each of these operations are equivalent to a relative phase shift between appropriate choices of polarization axes. However, it is convenient to consider them separately here for clarity.

Using this set of manipulations, we show in Fig.~\ref{fig:scheme} how to perform arbitrary single-qubit operations. The operation uses a polarization coupling, equivalent to a variable beam-splitter between the two polarizations, and two relative phase shifts applied to one polarization. It is well known that this is sufficient for local operations on a single qubit~\cite{Simon1990}. 

\begin{figure}[htbp]
\centering
\includegraphics[width=8cm]{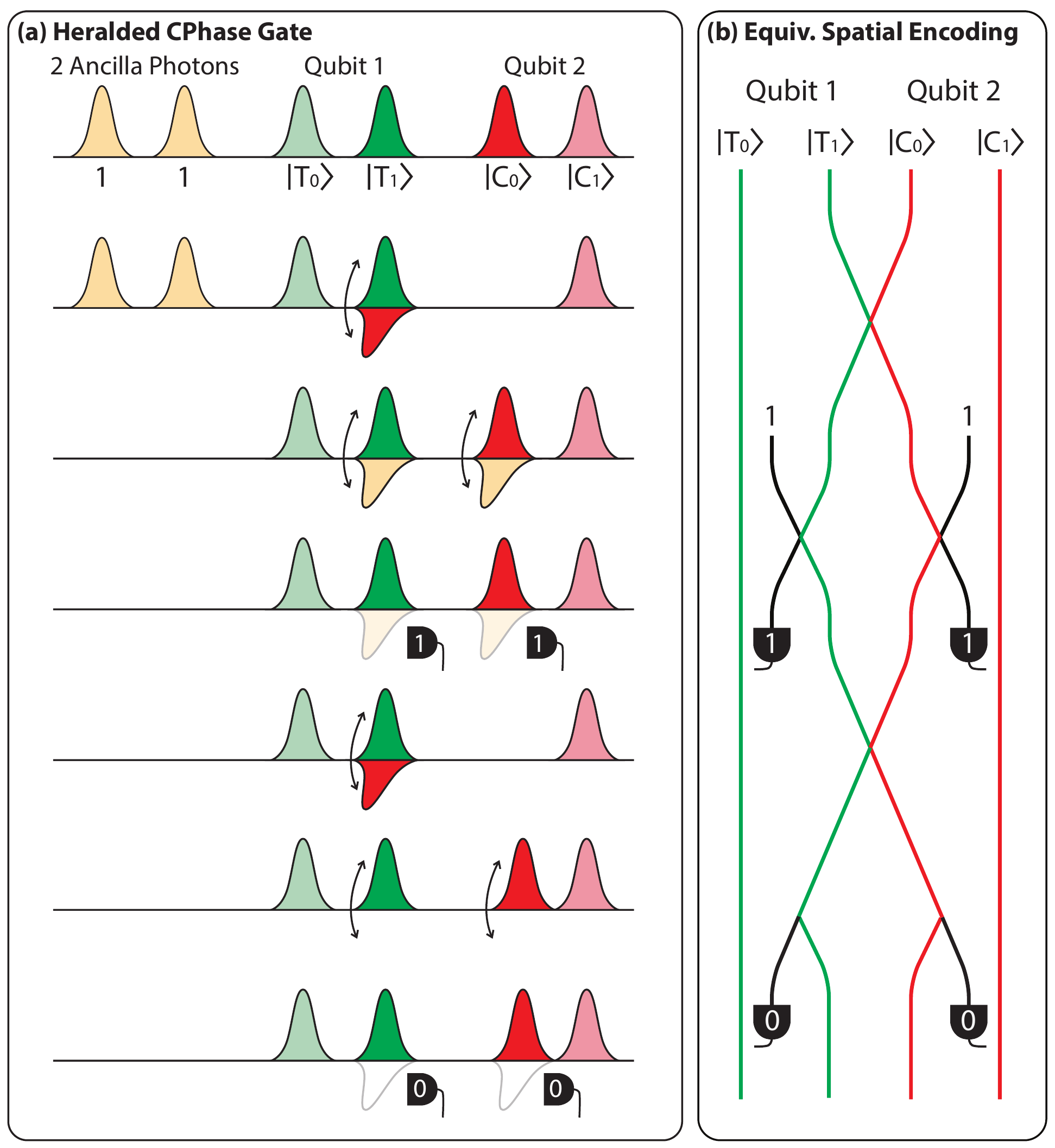}
\caption{(a) Scheme for heralded KLM CPhase gate using two ancilla photons. Note that these ancilla photons are not encoded as qubits, and each occupies a single time bin. Displacement and rotation operations are omitted for brevity. The numbers on detectors represent the number of photons detected in order to herald successful gate operation. This set of operations could be enacted by four sets of the elements shown in Fig.~\ref{fig:scheme}(c), along with appropriate read out elements. (b) Equivalent spatial scheme. Lone numbers represent input ancilla photons.}
\label{fig:CPhasescheme}
\end{figure}

In Fig.~\ref{fig:CPhasescheme}, we provide a sequence of operations to perform a time-bin heralded KLM-CPhase gate~\cite{Ralph2001,Pittman2001a} using two ancilla photons, sufficient to realize the entire KLM scheme in combination with single-qubit operations~\cite{Knill2001}. This can be trivially combined with local operations to perform a heralded controlled-NOT gate. The proposed scheme could be implemented using four of the sets of the elements in Fig.~\ref{fig:scheme}. Alternatively, since each stage of the operation returns the qubits to a single mode and polarization, the string could simply be sent through the same processing elements four times. In this way, the simple set of elements shown could be used to enact arbitrary multi-gate operations. We observe that our scheme is equally relevant to cluster state computing~\cite{Kok2007}, as it also allows the implementation of type-I and type-II fusion operations~\footnote{See the supplementary material for implementations of fusion gates in our scheme}, suggesting that its utility may extend beyond circuit based quantum computing protocols.

Our scheme is well suited to exploit the readily accessible high dimensionality and robustness of time-frequency modes to environmental dephasing noise. Fast switchable elements can enact different transformations on multiple time bins in a single pass, potentially allowing a substantial reduction in the required number of physical circuit elements. These advantages suggest that this scheme would naturally complement near-deterministic single photon sources~\cite{Nisbet-jones2013, Lindner2009}, for which significant challenges exist in building many identical sources. In this case, a single repetitive source would prepare the computational resource state: a string of otherwise indistinguishable single photons in pure quantum states consisting of multiple time bins of a single spatial mode. Our scheme then circumvents the complexity and spatial requirements involved in converting many temporally encoded photons into a spatial encoding. Further, combining temporal and spatial degrees of freedom may enable a significant increase in information capacity~\cite{Barreiro2008}.

\emph{Implementation-} 
Here we elaborate on a specific practical implementation of our scheme and discuss its feasibility within the current state of the art. As mentioned above, the basic logical operations are equivalent to a relative optical path length difference between a suitable choice of polarization axes. The appropriate experimental approach to generating these path length differences will depend on the specific time-bin structure that is used, as the bandwidth of different photon sources, and thus the time-bin duration, can differ by several orders of magnitude~\cite{Eisaman2011}. Here we will consider time bins with sub-ps duration; these are suitable for heralded single photons from spontaneous parametric down-conversion pumped by a pulsed laser. Bin-to-bin delay is set by the pump-pulse repetition period, which has been reduced below 10 ps in a number of systems~\cite{Oehler2010,Dromey2007,Wilcox:12}.

Polarization rotation and polarization coupling operations require a programmable birefringent element that independently manipulates each time bin. The switching time for this element must be less than the delay between consecutive time bins. A suitable integrated optical switch based on cross-phase modulation in a fiber has demonstrated a switching window of 10 ps~\cite{Hall2011}. As cross-phase modulation is polarization sensitive, this technique could be adapted to create fast-switched birefringent elements. 

The detector time resolution does not constrain the bin-to-bin delay since switching allows arbitrary time-bin components to be moved to the processing polarization or even to a separate read-out spatial mode for detection. Therefore read-out can be achieved with standard photon-number resolving detectors, including spatially~\cite{Jahanmirinejad2012,Divochiy2008} and temporally-multiplexed~\cite{Fitch2003,Achilles2003} single-photon detectors as well as transition edge sensors~\cite{Gerrits2011a,Lamas-Linares2012}.

For a displacement operation, a simple approach is to use a birefringent element that effects a polarization-dependent path length difference equal to integer multiples of the time-bin separation. A few-cm length of calcite would achieve a displacement of 10 ps. Alternatively, a delay loop could be used, coupled to the main mode by a Mach-Zehnder interferometer. A \(\pi\) phase shift created in this interferometer for only one polarization would couple that polarization into the delay line. The controllable phase shift could then be set to keep this polarization in the delay loop for an arbitrary integer number of loops, delaying it with respect to the primary set of time bins. A 3 mm delay line, possibly implemented on an integrated photonic chip, would create a 10 ps displacement. Although the scheme is no longer entirely single spatial mode with the use of a delay line, the arbitrary number of delay steps it allows may be desirable for faster processing. Finally, in the near future, it should be possible to use a quantum memory to reorder time bins arbitrarily, as demonstrated with classical pulses in a warm-vapor gradient echo memory~\cite{Hosseini2009}. This could provide a significant reduction in the number of individual operations needed.\\

\begin{figure}[htbp]
\centering
\advance\leftskip-0.5cm
\includegraphics[width=8cm]{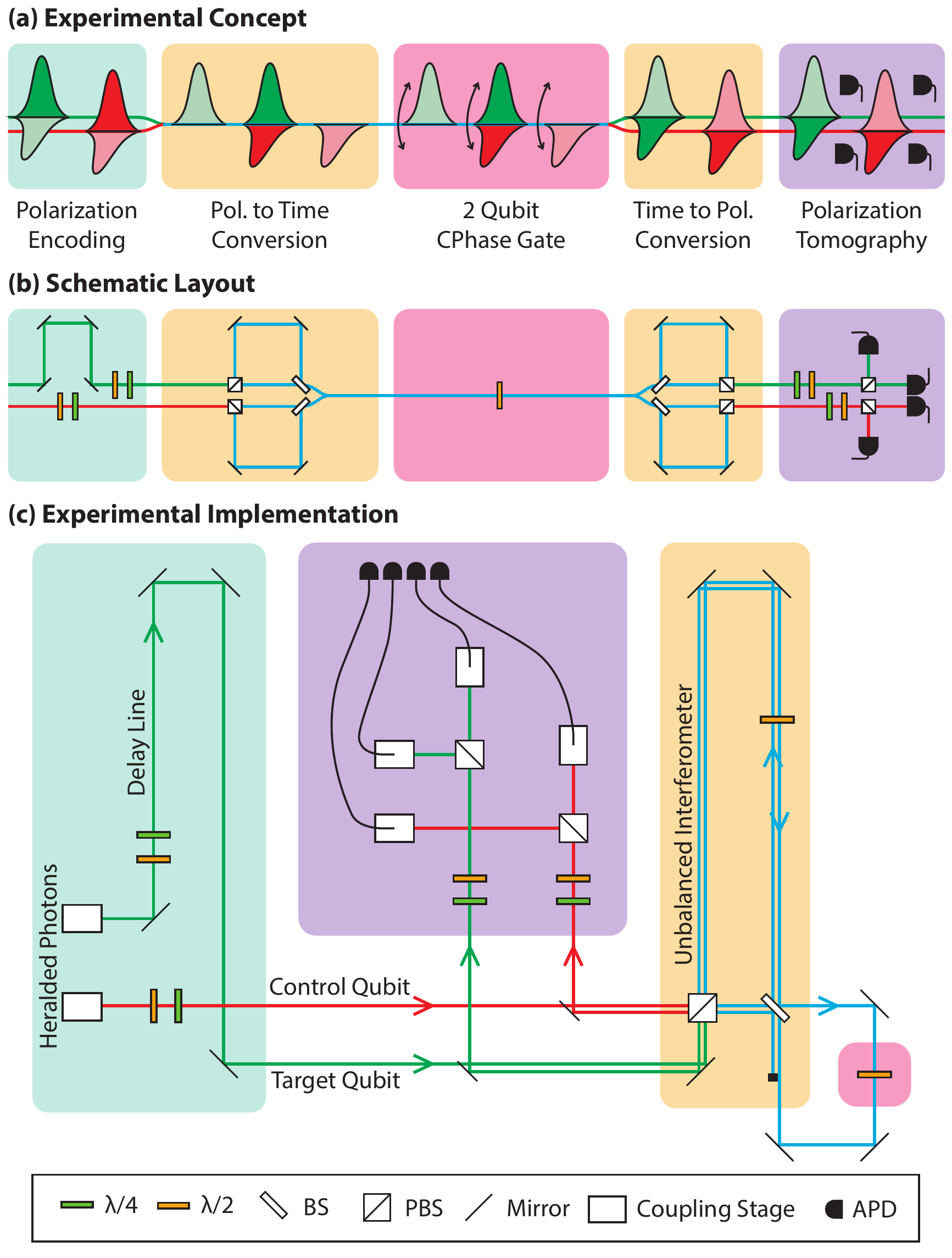}
\caption{(a) Concept for a single-spatial-mode CPhase gate with preceding state-preparation and following measurement stages. The photons are spectrally degenerate, and are color coded here for clarity. (b) Schematic of the associated experimental layout. Waveplates are used to encode polarization states for both the target photon (green) and the control photon (red). The target photon is delayed with respect to the control photon, and both are coupled into unbalanced interferometers for conversion of polarization encoding to time encoding. The photons are then combined into a single spatial mode in which a two-qubit gate is implemented using a half-wave plate. Conversion back to polarization-encoding states again uses unbalanced interferometers. Finally, polarization tomography is carried out using four avalanche-photodiode (APD) detectors. (c) Actual experimental implementation. Two SPDC sources provide heralded single photons for the experiment, which proceeds as described above, except that a single unbalanced interferometer is used instead of the four separate unbalanced interferometers for conversion between polarization-encoding and time-encoding.}
\label{fig:layout}
\end{figure}

\emph{Experiment-} In order to demonstrate the feasibility of our scheme, we have built an entirely single-spatial-mode post-selected CPhase gate for time-encoded qubits~\footnote{See the supplementary material for further details on the experiment}. Our gate is equivalent in principle to previous implementations~\cite{Ralph2002, O'Brien2003} that use spatial encoding, often along with a second degree of freedom such as polarization. Preceding the gate is a polarization-to-time conversion stage, and following it a time-to-polarization conversion stage allows for measurement. The experimental layout is shown in Fig.~(\ref{fig:layout}). At the core of our experiment, a single-spatial-mode gate is enacted. In this proof-of-principle experiment, we have replaced birefringent switches with passive beam splitters and a second spatial mode, as this allows us to readily incorporate two-mode analogues of single-spatial-mode single-qubit rotations and displacement operations (Fig.~\ref{fig:scheme}).

Two SPDC pair sources are used to provide two heralded pure single photons~\cite{Mosley2008}. Initially one qubit is encoded in the polarization of each photon. The qubits are then converted to a time basis using an unbalanced interferometer, producing two orthogonally polarized photons in a common spatial mode. One of the photons is delayed so that its first time bin coincides with the second time bin of the other photon. The gate operation is implemented by using a half waveplate to couple the polarizations, resulting in Hong-Ou-Mandel interference~\cite{Hong1987} between the two coincident time bins and allowing a post-selected non-linear interaction.

\begin{figure}[htbp]
\begin{center}
\includegraphics[width=6cm]{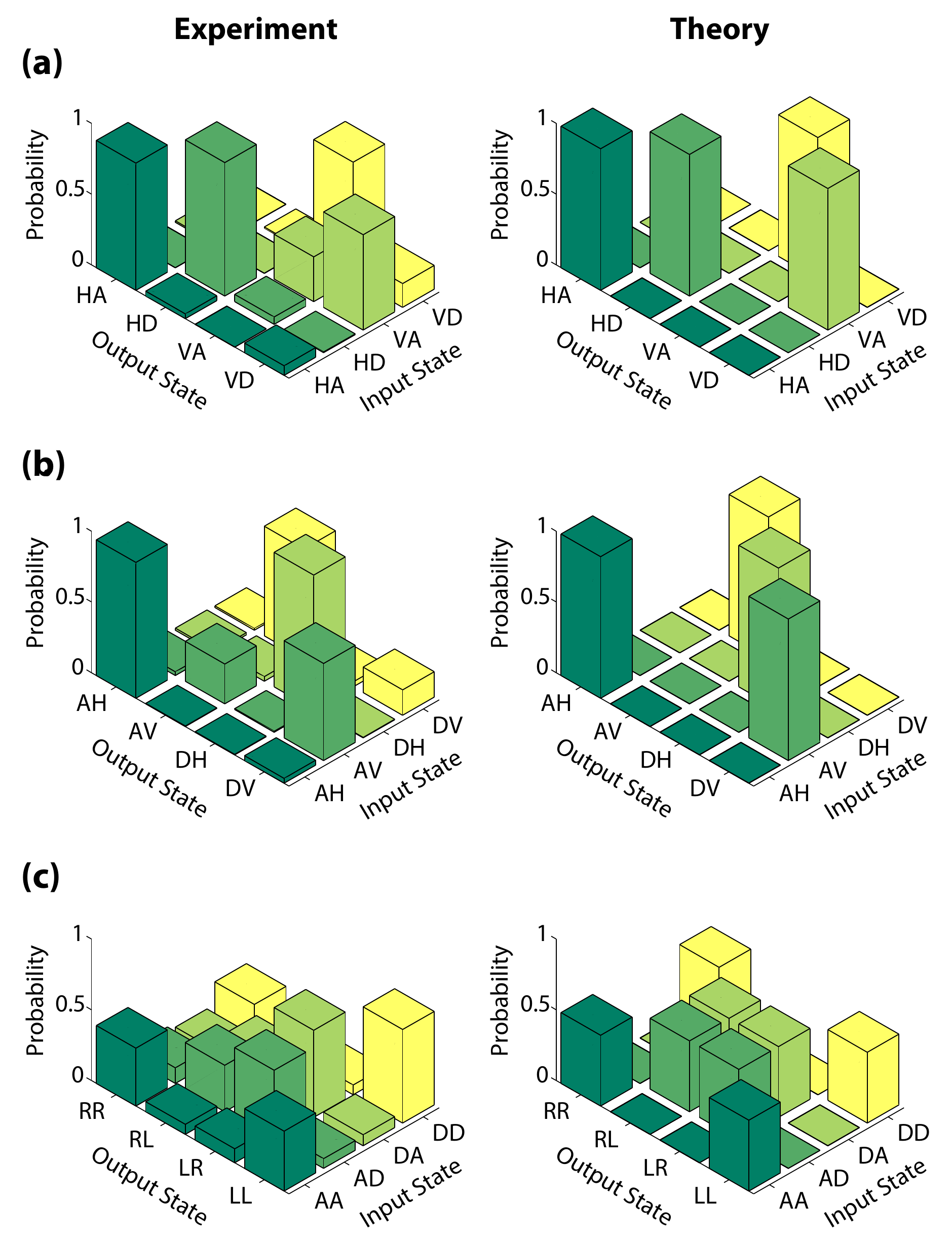}
\caption{Two-qubit output state measurements: (a) Input $H$-$V$ and $A$-$D$ bases for control and target photons respectively, output $H$-$V$ and $A$-$D$ bases. The measured classical fidelity for this operation is $F_{HA} = 0.84\pm0.03$. (b) Input $A$-$D$ and $H$-$V$ bases, output $A$-$D$ and $H$-$V$ bases, resulting in $F_{AH} = 0.84\pm0.02$. (c) Input $A$-$D$ bases for both photons, output $R$-$L$ bases for both photons, resulting in $F_{AA} = 0.85 \pm 0.06$. Theoretical ideal outputs are shown for comparison.}
\label{fig:TwoQubit}
\end{center}
\end{figure}

To characterize the two-photon operation of our gate, we initially input a control photon with a horizontal ($H$) or vertical ($V$) polarization and a target photon with an anti-diagonal ($A$) or diagonal ($D$) polarization. For these inputs, the CPhase gate should swap the target photon polarization between $A$ and $D$ if the control photon is $V$ polarized. The measured gate outcomes are shown in Fig.~(\ref{fig:TwoQubit}), where the control and target photons are measured in the $H$-$V$ and $A$-$D$ bases, respectively. For these bases, we define a classical fidelity measure~\cite{Hofmann2005} 
\begin{align}
F_{HA} = 1/4 [P(HA | HA) + P(HD | HD)\notag\\
 + P(VD | VA) + P(VA | VD) ]
\end{align}
where, for example, $P(VA | VD)$ represents the conditional probability of measuring outputs $V$ and $A$ given input $V$ and $D$ for the control and target photons respectively. We measure a classical fidelity of $F_{HA} = 0.84\pm0.03$. Changing the photon inputs to the control $A$-$D$ and target $H$-$V$ bases and also measuring in these bases, equivalent to transforming the bases by a Hadamard operation, allows us to measure a complementary fidelity $F_{AH}$. For this latter case, we measure a similar fidelity $F_{AH} = 0.84\pm0.02$. 

Following~\cite{Hofmann2005}, we use these fidelity measures to bound the quantum process fidelity. The resulting bound of the gate process fidelity $F_\mathrm{process}$
\begin{equation}
F_{AH} + F_{HA} - 1 \leq F_\mathrm{process} \leq \mathrm{Min}[F_{AH}, F_{HA}]
\end{equation}
is calculated to be $0.68 \pm 0.04 \leq F_\mathrm{process} \leq 0.84 \pm 0.02$, comparable to other optical two-qubit gate implementations \cite{O'Brien2003, Langford2005}.

An alternative measure of our gate fidelity demonstrates its non-classical operation. For this, we consider an additional choice of bases with both inputs in the $A$-$D$ basis, and both outputs in the $R$-$L$ (right-left) basis. We measure the classical fidelity for this operation to be $F_{AA} = 0.85 \pm 0.06$. As shown in~\cite{Hofmann2005a}, since this measure, along with $F_{HA}$ and $F_{AH}$, are all greater than 2/3 the gate operation must be non-classical. Our gate exceeds this criterion with 99.8\% confidence.

The gate fidelity is limited by the spatial mode overlap of our photons. Due to the long path length in the time-to-polarization converter, this overlap is sensitive to the slight changes in alignment caused by temperature variations and vibrations. This path length is necessary to achieve a delay between consecutive time bins  that is resolvable by the coincidence counting electronics and detectors~\footnote{See the supplementary material for further details on the experiment}. We modeled this effect by calculating the ideal gate operation on partially distinguishable input photons in the states $\ket{\psi}$ and $\alpha \ket{\psi} + \sqrt{1-\alpha^2} \ket{\psi_\mathrm{disting.}}$ respectively, and found that $\alpha = 0.91$ minimized the L1 distance between the results and theoretical predictions. \\

\emph{Conclusions-} We have presented a scheme for linear optical quantum computing using time-bin encoded qubits in a single spatial mode. We have shown how to implement arbitrary single-qubit operations and a heralded CPhase gate as required for universal quantum computing in the KLM scheme. In support of this concept, we have demonstrated a novel post-selected single-spatial-mode two-qubit CPhase gate. We measured an average classical gate fidelity of $0.84 \pm 0.07$ across 3 different bases, confirming its non-classical operation. 

We thank J. Nunn for helpful discussions. This work was supported by the Engineering and Physical Sciences Research Council (EP/H03031X/1), the European Commission project Q-ESSENCE (248095) and the Air Force Office of Scientific Research (European Office of Aerospace Research and Development).

\bibliography{TimeBinPapers.bib}

\section{Supplementary Information}
\begin{figure}[htbp]
\begin{center}
\includegraphics[width=8.5cm]{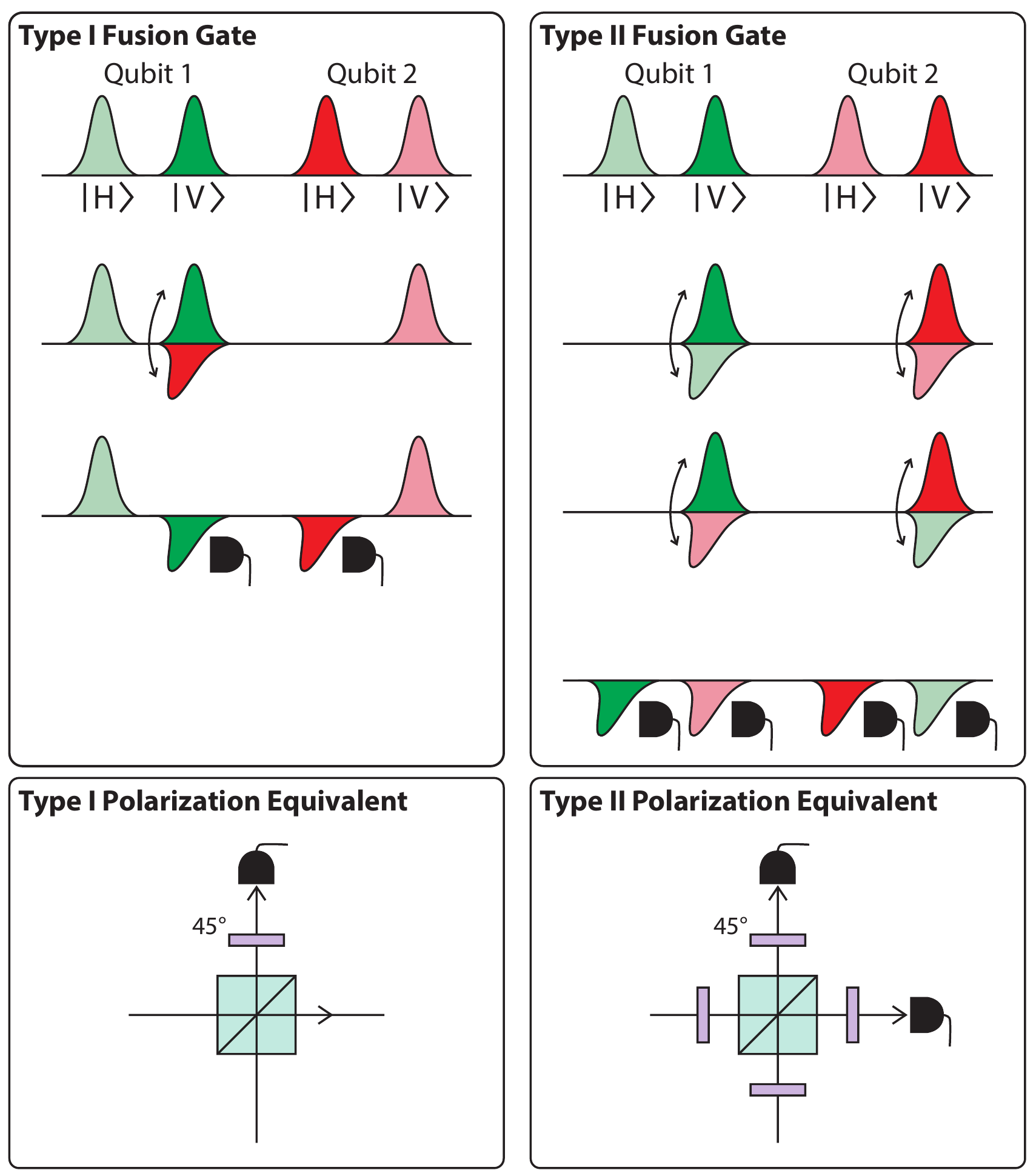}
\caption{Protocols for the implementation of type-I and type-II fusion operations, with equivalent spatial analogues.}
\label{fig:cluster}
\end{center}
\end{figure}

\emph{Fusion gates-} In the main text we outline methods for universal linear-optical quantum computing (LOQC) using time-bin-encoded qubits and the Knill-Laflamme-Milburn ~\cite{Knill2001} scheme. In Fig.~\ref{fig:cluster} we additionally provide protocols for the  type-I and type-II fusion gates necessary for many cluster state based quantum computing schemes~\cite{Kok2007}.

\emph{Experiment-} Our experiment uses two spontaneous parametric down-conversion sources for the generation of heralded single photons. An 80MHz Ti:Sapphire oscillator (Mai-Tai, Spectra Physics) producing 100 fs pulses at 830nm (2.6W average power) is up-converted to 700mW of 415nm light via a 700$\mu m$  BaB2O4 (BBO) crystal cut for type-I second-harmonic generation. This is split on a 50:50 beam splitter and used to pump two 8mm-long AR-coated Potassium Dihydrogen Phosphate (KDP) crystals phase-matched for degenerate type-II collinear parametric down-conversion. We spend time optimizing the collection optics and spatial mode-matching to achieve a coincidence count rate of 160kHz on each crystal with a raw heralding efficiency of 28-30\% without any filters. The source is designed to be spectrally factorable~\cite{Mosley2008} which improves the heralding efficiency we can achieve when interference filters (Semrock, $\Delta\lambda = 3$nm) are used to match the bandwidths of the broad and narrowband daughter photons. With the filters in place we achieve a four photon coincidence rate of 20 Hz when measured directly from the sources.

Two heralded single photons from these sources are initially used to encode qubits in the polarization state of each photon using $\lambda/2$ and $\lambda/4$ waveplates. To provide a concise mathematical description of our photons, we will label them the `control' and `target' photon respectively. 
\begin{align}
\text{Control Photon: } \ket{\psi_C} = \alpha_C \ket{H} + \beta_C \ket{V}\notag\\
\text{Target Photon: } \ket{\psi_T} = \alpha_T \ket{H} + \beta_T \ket{V}
\end{align}

The target photon is delayed with respect to the control photon, and then both are coupled into the same polarizing beam splitter (PBS). This is used to send the polarization components down different arms of a 1.5m unbalanced interferometer. Bringing the two components back together using a balanced beam splitter finishes the polarization to time-encoding conversion. The final beam splitter can only recombine the time-bin components probabilistically, although the failure modes come out of the wrong port of the beam splitter, and so do not contaminate the rest of the experiment. This could be replaced by an active switching element to deterministically recombine the time bins into a single spatial mode. 

After this conversion, the two photons are orthogonally polarized along a common mode, with the target photon delayed so that its first time bin coincides with the second time bin of the control photon. Due to the interferometric technique used for polarization to time conversion, relative phases are acquired by different qubit components. These are denoted by $\theta_{C1}$ and $\theta_{T1}$, where we have used the convention that the phase is applied to the delayed component. The qubits are now in the state
\begin{align}
\ket{\psi_C} = \alpha_C \ket{1H} + e^{\mathrm{i} \theta_{C1}} \beta_C \ket{2H} \notag\\
\ket{\psi_T} = e^{\mathrm{i} \theta_{T1}} \alpha_T \ket{3V} + \beta_T \ket{2V}
\end{align}
where, for example, \(\ket{1H}\) denotes a photon in time-bin 1 and polarization H, and the time bins are numbered sequentially from earliest to latest. This encoding allows the gate operation to be implemented using a single half waveplate as a variable beam splitter between the two polarizations, creating Hong-Ou-Mandel interference between the two coincident time bins. When the axes of the waveplate are aligned with photon polarizations, the gate operates with identity, while at 27.4 degrees, it implements a CPhase operation. The non-overlapped time bins also couple with ancillary loss modes due to this polarization beam splitter, analogously to the coupling to spatial ancilla modes in a more conventional CNOT gate. As with other implementations of this scheme, the gate only succeeds with probability 1/9. A successful operation maps
\begin{align}
&\alpha_1 \ket{1H}\ket{3V}  + \alpha_2 \ket{1H} \ket{2V} + \alpha_3 \ket{2H} \ket{3V} \dots\notag\\
&\quad+ \alpha_4 \ket{2H} \ket{2V}\notag\\
\rightarrow \quad &\alpha_1 \ket{1H}\ket{3V} + \alpha_2 \ket{1H} \ket{2V} + \alpha_3 \ket{2H} \ket{3V} \dots\notag\\
&\quad- \alpha_4 \ket{2H} \ket{2V}\\
&\text{where } \alpha_1 =e^{\mathrm{i} \theta_{T1}}  \alpha_C \alpha_T, \; \alpha_2 =  \alpha_C \beta_T,\notag\\
& \quad \alpha_3 = e^{\mathrm{i} (\theta_{T1}+\theta_{C1})}  \alpha_T \beta_C, \; \alpha_4 = e^{\mathrm{i} \theta_{C1}}  \beta_T \beta_C\notag
\end{align}

After the gate, the photons are re-injected into the same unbalanced interferometer in the other direction (with the components in the long arm again gaining relative phase terms, this time denoted $\theta_{C2}$ and $\theta_{T2}$). This allows the time-bin encoding to be decoded back into polarisation, after which polarisation tomography can be carried out to measure the state of the qubits. 
\begin{align}
\ket{\psi_C} =  e^{\mathrm{i} \theta_{C2}} (\alpha_C \ket{2V} + e^{\mathrm{i} (\theta_{C1}-\theta_{C2})} \beta_C \ket{2H})\notag\\
\ket{\psi_T} =  e^{\mathrm{i} \theta_{T1}} (\alpha_T \ket{3V} + e^{\mathrm{i} (\theta_{T2}-\theta_{T1})} \beta_T \ket{3H})\label{eqn:decoding}
\end{align}

The reuse of the initial encoding interferometer creates an intrinsically phase stable encoding and decoding. As can be seen in Eqn.~(\ref{eqn:decoding}), if $\theta_{C1}-\theta_{C1}$ and $\theta_{T2}-\theta_{T1}$ are constant, the operation will be unaffected. This removes the need for phase stabilisation, although slow drifts in the alignment of the paths must be corrected for in order to ensure that the encoding is kept the same. This was accomplished by using a $\lambda/4, \lambda/2, \lambda/4$ series of waveplates in the output paths to correct for the relative phase between the horizontal and vertical polarisations. Before each basis set measurement, the half waveplate was adjusted to maximise the decoding fidelity for input diagonally polarised photons when measured in the diagonally polarised basis.

The output photons were detected using an array of four avalanche photodiode (APD) single photon counting modules (PerkinElmer SPCM-AQ4C). Due to the loss modes and non-deterministic decoding of the photons, the specific time bins of the output qubits must be measured separately. Therefore the outputs from the APD modules were each split into four different channels with different temporal delays, and monitored
by a home-built coincidence counting program loaded onto a commercially available FPGA development board (Xilinx SP605) operating with a 2.86 ns coincidence window. The resulting set of 16 signals (and two herald signals) covers the 4 time bins for each spatial mode and polarisation, allowing the qubit state to be reconstructed.
\begin{figure}[htbp]
\begin{center}
\includegraphics[width=7cm]{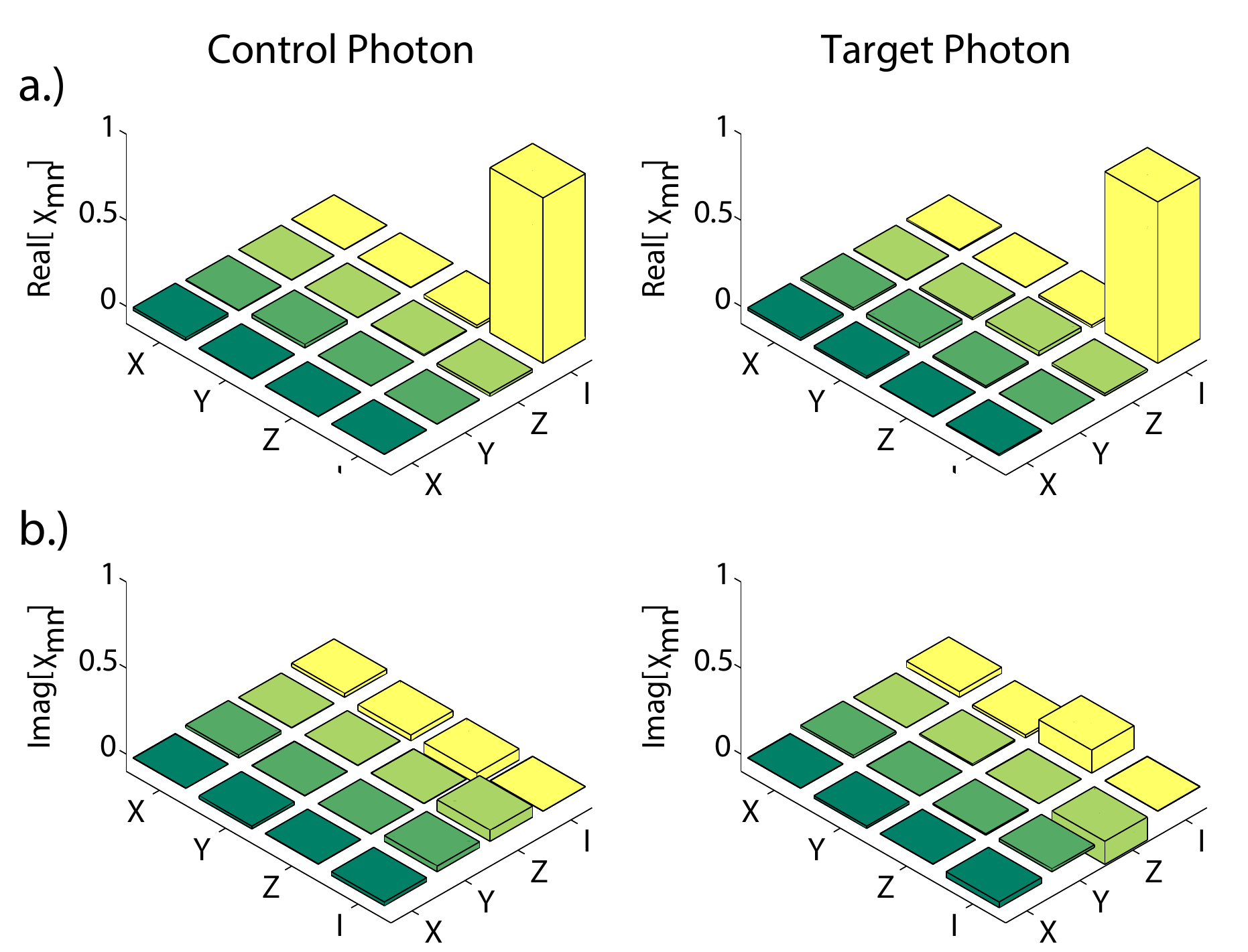}
\caption{Single photon encoding and decoding performance. (a) Real and (b) imaginary parts of single qubit polarisation state tomography data for the `control' and `target' input photons respectively. Each shows a high fidelity with the identity.}
\label{fig:SingleQubit}
\end{center}
\end{figure}

In Fig.~(\ref{fig:SingleQubit}) we present data showing the high fidelity of the polarisation to time conversion. For the control and target photons, process tomography for the mapping from input polarisation state to output polarisation state gives a fidelity with the identity of $0.960\pm0.001$ and $0.936\pm0.001$ respectively. This shows that we can reliably create time-bin encoded qubits, and maintain their coherence across the setup. The error in the fidelity is due to the slight deviation of the non-polarising beam splitter away from its ideal reflectivity, and due to differences in coupling and loss between the different time-bins.

\end{document}